\documentclass[aps,a4paper,twocolumn,superscriptaddress,preprint,showpacs,showkeys,10pt]{revtex4}
\usepackage{graphicx}
\usepackage{subfigure}
\usepackage{latexsym}   
\usepackage{amsmath}    

\begin{document}

\title{Extended gaussian ensemble solution and tricritical points of a system with long-range interactions}
\author{Rafael B. Frigori}
\email{frigori@utfpr.edu.br} 
\affiliation{Universidade Tecnol\'ogica Federal do Paran\'a, \\
             Rua XV de Novembro 2191, CEP 85902-040 Toledo, PR, Brazil.}
\author{Leandro G. Rizzi}
\email{lerizzi@usp.br} 
\author{Nelson A. Alves}
\email{alves@ffclrp.usp.br} 
\affiliation{Departamento de F\'{\i}sica e Matem\'{a}tica, FFCLRP, \\
             Universidade de S\~ao Paulo,
             Avenida Bandeirantes, 3900 \\ 
             14040-901, Ribeir\~ao Preto, SP, Brazil.}

\begin{abstract}
   The gaussian ensemble and its extended version theoretically play the important role of interpolating ensembles between the microcanonical and the canonical ensembles.
   Here, the thermodynamic properties yielded by the extended gaussian ensemble (EGE) for the Blume-Capel (BC) model with infinite-range interactions are analyzed.
   This model presents different predictions for the first-order phase transition line according to the microcanonical and canonical ensembles.
   From the EGE approach, we explicitly work out the analytical microcanonical solution.
   Moreover, the general EGE solution allows one to illustrate in details how the stable microcanonical states are continuously recovered as the gaussian parameter $\gamma$ is increased.
   We found out that it is not necessary to take the theoretically expected limit 
 $\gamma \rightarrow \infty$ to recover the microcanonical states in the region between the canonical and microcanonical tricritical points of the phase diagram.
   By analyzing the entropy as a function of the magnetization we realize the existence of unaccessible magnetic states as the energy is lowered, leading to a breaking of ergodicity.
   
\end{abstract}


\maketitle

\section{Introduction}

  The canonical and grand-canonical ensembles approximate the 
microcanonical ensemble in the limit of infinitely large number of particles, where 
surface effects and fluctuations can be disregarded with respect to the bulk mean values
\cite{Gross_book,Gross_report}.
  However, if the system sizes are not large enough compared to the range of interactions, or even in the
presence of long-range forces, this inherent expectation changes dramatically.
  Although such nonextensive systems can be appropriately described by models in a volume-dependent   scaling manner \cite{Kac}, the non-additive character still remains.
  As a matter of fact, the lack of additivity can be noticed for astrophysical objects, where gravitational interaction is responsible for a nonnegligible contribution from particles at large distances \cite{pad_gravitational}.
  Thus, one realizes that most of the systems in nature can be encompassed in a class that can be designated non-additive for what concerns energy and entropy.
  The existence of a fundamental ensemble, the microcanonical one, 
seems to meet a consensus, while others, in particular the canonical and grand-canonical ones, are taken as its approximations
\cite{Gross_book,Gross_2000,Gross_2002}.
   In cases where full consistency of statistical ensembles holds for systems that undergo phase transitions, it is found that finite size scaling relations still place the microcanonical approach
as the fundamental one \cite{kastner_JSP2001}.

  There are many examples of systems whose equilibrium properties are not equivalent in both
microcanonical and canonical ensembles. 
  Differences in the thermodynamic features have been verified analytically for systems with long-range interactions \cite{Ruffo_PRL2001,Ruffo_PRL2005,Ruffo_JSP2005,Touchette_A2007}.
  These examples show that the nonequivalence appears where the canonical ensemble presents a phase diagram with a 
first-order transition line.
  Actually, necessary and sufficient conditions for equivalence of ensembles can be formally stated \cite{TurkingtoJSP101}.
  Thus, apart from the expected difference in the intermediate values of extensive thermodynamic quantities when one works with finite systems, the non-additive property also sets striking differences in the thermodynamic limit, leading to different phase diagrams \cite{kastner_PRL2006,kastenerJSP2006}. 
  Such nonequivalence has its counterpart in the nonconcavity of the entropy as a function of energy, $S=S(E)$  
 \cite{Ruffo_PRL2001,kastner_A2007,Bouchet_JSP118}. 
  This may result in uncommon features at first-order phase transitions like temperature discontinuity 
and negative specific heat in the microcanonical ensemble  
\cite{Ruffo_PRL2001,TurkingtoJSP101,kastner_A2007,Ruffo_EPJB2008,Ruffo_Lecture2002,Ellis_A335}.  
  In turn, the thermodynamic temperature  $\beta = 1/T(E) = \partial S/ \partial E$,  (we take the Boltzmann constant $k_B=1$) is not a monotonic function of the energy and the equilibrium value $E(\beta)$ may be a multivalued function of $\beta$  \cite{Gross_book,Ruffo_PRL2001,kastner_A2007}.
         
    An alternative ensemble, the gaussian ensemble \cite{Hether_JLTP66,Hether_PRD35,Hethe_87,Challa_PRL60,Challa_PRA38}
was introduced to deal with systems that exchange energy with a finite reservoir.
    This contrasts with the canonical ensemble, where the system is in thermal contact with a huge heat reservoir 
and the energy exchange is controlled by the temperature of the reservoir, which defines the average energy of the system.  
    On the other hand, in the limit of no energy exchange with the reservoir, the system is isolated and thus has fixed energy.
    This is the microcanonical point of view, whose experimental situation resembles a system in contact with a fictitious reservoir of extremely small size where the energy exchange can be disregarded.
     The gaussian ensemble has also been described as a regularization procedure for the microcanonical ensemble \cite{Lukkarinen_99}.    
    Later on, Johal {\it et al.} \cite{Vives_PRE68} redefine the assumptions characterizing the gaussian ensemble to describe the thermodynamic properties of a system also in contact with a finite reservoir.
    This led to an extended version of the former gaussian ensemble.
    The extended gaussian ensemble (EGE) presents a smooth interpolation  between its limiting behaviors, corresponding to the microcanonical and canonical ensembles.

    This work explores the EGE as a working ensemble for a system where the ensemble nonequivalence has been demonstrated, the Blume-Capel model.
    We explicitly work out the analytical microcanonical solution from this ensemble.
    By means of the general EGE solution we are able to illustrate in details how the stable microcanonical states are continuously recovered as the gaussian parameter $\gamma$ is increased.
    We investigate the EGE behavior in the region of the phase diagram where one observes a first-order phase transition line in the canonical ensemble but of second-order type in the microcanonical description.
    Then, we point out how EGE identifies the canonical and microcanonical tricritical points.
    Moreover, we call attention to the broken ergodicity found in this model.

    The EGE formulation encompasses a natural extension of Statistical Mechanics, to include non-additive  systems.
    The relation between EGE and Tsallis statistics has been described in Ref. \cite{Vives_PRE68,Morishita_07}, with the Tsallis parameter $q$ being related to the parameter $\gamma$ in the EGE.
    The theoretical background characterizing this ensemble is presented in Section 2,
where we briefly review some thermodynamic relations that are $\gamma$ dependent.
    The EGE solution of the mean-field BC model is carried out in Sec. 3 and 
is confronted at the thermodynamic level with the usual solutions 
in the canonical and microcanonical ensembles. 
    The main conclusions about the effectiveness of the EGE in determining thermodynamic properties
are summarized in Sec. 4.
   

\section{Extended gaussian ensemble}

  The canonical ensemble describes thermal properties of a system in thermal equilibrium with a heat reservoir.
  A new insight has been obtained when the reservoir is finite and possibly small.   
  To this end, let $a$ be a system with energy $E$ and entropy $S$, and $b$ a reservoir with energy $E_b$ and entropy $S_b$, which exchanges energy with $a$.
   As a consequence, the energy of the system is allowed to fluctuate.
   Both systems form an isolated system with total energy $E_t= E + E_b$ and total entropy $S_t$.
   Equilibrium is reached when the total entropy $S_t(E)$ is a maximum. 
   The system itself and its heat bath can be considered subsystems of an isolated system where
$E$ fluctuates around its mean value $U$.
   Thus, for fixed external parameters like total energy $E_t$ and number of particles in the system, the most probable energy $U$ is such that the expansion of the reservoir entropy $S_b$ around its equilibrium value $E_t-U$ can be written up to the second order as 
\begin{eqnarray}
 \lefteqn{ S_b(E_b) = S_b(E_t - U) +\left(\frac{dS_b}{dE_b}\right)_{E_t-U}(U -E)  } \notag  \\
      & & +      \frac{1}{2} \left(\frac{d^2S_b}{dE_b^2}\right)_{E_t -U} (U -E)^2 + \cdots
                       \, .                                               \label{eq:expansion}
\end{eqnarray}
   Because the derivatives depend on the reservoir thermodynamic properties, one defines
\cite{Vives_PRE68}
\begin{equation}
  \left(\frac{dS_b}{dE_b}\right)_{E_t-U}    =  \alpha  \, ,
\end{equation}
and
\begin{equation}
   \frac{1}{2} \left(\frac{d^2S_b}{dE_b^2}\right)_{E_t -U} =  - \gamma   \,  .  
\end{equation}
    In the case of an infinite reservoir, one would be working with the canonical ensemble and
$\alpha$ would thus be identified with the inverse thermodynamic temperature, $\alpha = 1/T$.
    This is because in the canonical ensemble approach the temperature $T$ of the reservoir is a fixed parameter that
determines the mean energy of the system.
    The effect of an infinite reservoir with constant temperature yields $\gamma =0$ and 
vanishing higher-order derivatives in Eq.(\ref{eq:expansion}); otherwise, those terms should be taken into account.

     The EGE is defined by the condition $\gamma \neq 0$ and probability density
\begin{equation}
  P_{\gamma,\alpha}(E) = \frac{\rho(E)\, e^{-\alpha E - \gamma (E- U)^2}}{Z_{\gamma}(U,\alpha)} \, ,
             \label{eq:prob}
\end{equation}
where $Z_{\gamma}(U,\alpha)$ stands for the normalization constant, which is the corresponding partition function in 
EGE \cite{Vives_PRE68,Touchette_JDP119,Touchette_E73},
with density of states $\rho(E)$ and parameters $\gamma$, $\alpha$, and the dependent one
$U= U(\alpha,\gamma)$.
   Actually, the extended gaussian ensemble is a particular case in a class
of general functions $g(E)$ \cite{Touchette_JDP119,Touchette_E73}; the quadratic form
$g(E)=\gamma(E-U)^2$ is just a convenient choice.
   The probability density in Eq. (\ref{eq:prob}) can be used to write the average energy of the system,
\begin{equation}
    U = \int E\, P_{\gamma,\alpha}(E) dE   \, .   \label{eq:U_self} 
\end{equation}

   Let us also introduce the extended thermodynamic potential analogous to the one
in the canonical approach, $ \Phi_{\gamma}(U,\alpha) = - {\rm ln}\, Z_{\gamma}(U,\alpha)$.
   From here, the derivative at fixed value $\gamma$ can be obtained,
\begin{equation}
    \left(\frac{\partial \Phi_{\gamma}}{\partial \alpha} \right)_{\gamma}  = U  \,  , \label{eq:U}
\end{equation}
which parallels that of the usual canonical approach.
   The average energy $U$ can be found self-consistently by means of Eq. (\ref{eq:U_self}), which recovers the usual canonical ensemble result for $\gamma=0$, or from Eq. (\ref{eq:U}), as describing the equilibrium average energy with fixed parameters $\gamma$ and $ \alpha$.  
   In this paper we follow a kind of inverse problem, $U$ will be set as an input parameter that must, in conjunction with the variational problem of minimization of the extended thermodynamic potential, satisfy stability conditions for some (unknown) temperature $1/\alpha$, which is $U$ dependent. 
   The extended heat capacity has also been introduced \cite{Challa_PRL60,Vives_PRE68},
$ C_{\gamma} = - \alpha^2 ({\partial U}/{\partial \alpha})_{\gamma}$.

   The usual canonical ensemble deals with homogeneous configurations in equilibrium as a function of
intensive variables like temperature.
   The canonical averages always produce smooth distributions of mean values, as in the case of heat capacity, when
averaged over fluctuations.
   In  contrast to the canonical heat capacity, the extended heat capacity may present negative values when $\gamma > 0$.
   Negative values of $C_{\gamma}(U)$ require that $ (\partial U/\partial \alpha)_{\gamma} > 0$.
   Thus, van der Waals loops can be seen in this formalism.
   This sort of behavior has been observed in typical caloric curves, temperature versus mean energies, for systems with thermodynamic first-order phase transition, a forbidden phenomenon in the canonical picture.    
   These features are illustrated in Fig. 2 for the Blume-Capel model.
   Thus, the standard homogeneous thermodynamics given by the canonical ensemble is not suited to describe first-order phase transitions. 
   On the other hand, the stability condition of a system is related to the homogeneous temperature that 
defines thermal equilibrium with the huge reservoir.
   The EGE includes the possibility of a rather small heat bath, thus allowing for the appearance of inhomogeneous configurations in the system for finite $\gamma$, which results into a weakened version for the constraint of constant energy that defines the microcanonical ensemble.

   The extended entropy can be obtained by the Legendre-Fenchel (LF) transform of the extended canonical 
thermodynamic potential $\Phi_{\gamma}(\alpha)$ as \cite{Touchette_JDP119,Touchette_E73,Vives_PRE68}
 \begin{equation}
  S_{\gamma}(U) =    \alpha \left(\frac{\partial \Phi_{\gamma}}{\partial \alpha}\right)_{\gamma}
                         + \gamma  \left(\frac{\partial \Phi_{\gamma}}{\partial \gamma}\right)_{\alpha} 
                                                      - \Phi_{\gamma}  \, .                          \label{eq:LF}
\end{equation}  
   From this transform and Eq. (\ref{eq:U}), it follows that $\alpha(U) = \partial S_{\gamma}/ \partial U$. 
   
   Notice that the above relations recover the canonical results in the limit $\gamma \rightarrow 0$.
   In this case, one has the standard Legendre transform
$  S(U) = \beta U - \Phi(\beta)$,
where $\Phi(\beta) ={\rm lim}_{\gamma \rightarrow 0} \Phi_{\gamma}(\alpha=\beta)$ corresponds to the canonical potential and $U$ is the equilibrium mean energy,  $U = \partial \Phi(\beta)/\partial \beta $.
  It is well known that the standard Legendre transform of $\Phi(\beta)$ always produces a concave function of $U$.
  Therefore, nonequivalence between microcanonical and canonical ensembles appears when the microcanonical entropy is a nonconcave function of $U$ in some energy range, as shown in Fig. 2b. 
  In that case of nonequivalence, $S(U)$ can be named as just the canonical entropy:
$ S_{\rm can} = \beta U_{\rm can}(\beta)  + {\rm ln}\,Z_{\rm can}(\beta) $.
  On the other hand, the limit $\gamma \rightarrow \infty$ corresponds to the microcanonical case.
  This can be seen as 
$ {\rm lim}_{\gamma \rightarrow \infty} \sqrt{\pi/\gamma} Z_{\gamma}(U,\alpha) $
through the use of the Dirac's delta sequence in the gaussian form \cite{Lukkarinen_99}.  
  For finite $\gamma$ one obtains an intermediary thermal description between the known limiting
ensembles.


\section{Extended gaussian solution of the mean-field BC model}

  The Blume-Capel model is a spin-$1$ Ising model \cite{blume_1966,capel_all}
and was introduced to describe phase separation in magnetic systems.
  It is a particular case of the Blume-Emery-Griffiths model \cite{beg_1971}
aimed at describing the critical behavior of He$^3$-He$^4$ mixtures with different concentrations.
  Here we consider its mean field version,
\begin{equation}
  H(S)= \Delta\sum_{i=1}^{N}S_{i}^{2} -\frac{J}{2N}\left(\sum_{i=1}^{N}S_{i}\right)^{2}   \, , \label{eq:H(S)}
\end{equation}
where $S_i = \{0,\pm 1 \}$.
  The couplings  $J> 0$ and $\Delta$ are the exchange and crystal-field interactions,
respectively.
  The BC model represents a simple generalization of the spin-$1/2$ Ising model, 
but with a rich phase diagram in the $(\Delta/J, T/J)$ plane.
  It exhibits a first-order transition line, tricritical point, and a second-order transition line.  
  
  The critical properties of the BC model can be determined analytically in both the microcanonical \cite{Ruffo_PRL2001}
and canonical ensembles \cite{beg_1971,beg_1974}.
  It has been demonstrated that these ensembles do not yield the same phase diagram for the first-order 
critical line \cite{Ruffo_PRL2001}.
  The canonical tricritical point occurs at $(\Delta/J, T/J) =(\simeq 0.46209812, 1/3)$,
which gives origin to the first-order transition line for larger values of $\Delta/J$. 
  The microcanonical solution identifies the tricritical point at  $(\Delta/J, T/J) \simeq (0.46240788,0.33034383)$.

  Here, it is useful to introduce the order parameters magnetization $M=\sum_{i=1}^{N} S_i= N_+ - N_{-}$ and 
its second moment, the quadrupole moment $Q= \sum_{i=1}^{N}  S_i^2 = N_+ + N_{-} $, 
where $N_+$ and  $N_{-}$ are, respectively, the number of sites with up and down spins.
  If $N_0$ is defined as the total number of zero spins, then
$N = N_+ + N_{-} + N_0 $ is the total number of spins in the system.

  The extended gaussian partition function  
\begin{equation}
    Z_{\gamma}(U,\alpha) = \sum_{\{S\}}e^{-\alpha H(S)-\gamma [H(S)-U]^2} \, ,      \label{eq:Z_BC}
\end{equation}
can be analytically solved in terms of its order parameters $M$ and $Q$.
    To this end, the so-called Hubbard-Stratonovich (HS) transformation
\begin{equation}
    e^{-bx^2/2}=\frac{1}{\sqrt{2\pi b}}\intop_{-\infty}^{+\infty} dy\, e^{-y^2/2b -ixy} \, ,
\end{equation}
is applied to the gaussian term in Eq. (\ref{eq:Z_BC}) with the choices $b=2\gamma$ and $x=H(S)- U$.
  It turns out that
\begin{eqnarray}
  \lefteqn{ Z_{\gamma}(U,\alpha) =  \sum_{\{S\}}\frac{1}{\sqrt{4\pi \gamma}}e^{-\alpha U} } \\
& & \times   \intop_{-\infty}^{+\infty}dy \,  
       e^{-\frac{y^2}{4\gamma}-(iy+\alpha) (\Delta\sum_{i=1}^{N} S_{i}^{2}-U) }\, 
       e^{(iy+\alpha) (\sqrt{\frac{J}{2N}}\sum_{i=1}^{N}S_{i})^2 }.  \notag   \\
&  &        \label{eq:ZG_BC}   
\end{eqnarray}

By making use of another HS transformation, the extended gaussian-partition function becomes
\begin{eqnarray}
  \lefteqn{ Z_{\gamma}(U,\alpha) = \frac{1}{\sqrt{4\pi \gamma}}\sum_{\{S\}}e^{-\alpha U}} \notag  \\
  & & \times  \,
     \intop_{-\infty}^{+\infty}dy\,   \left(\frac{iy+\alpha}{\pi}\right)^{1/2}
      e^{-\frac{y^{2}}{4\gamma}-(iy+\alpha)(\Delta\sum_{i=1}^{N}S_{i}^{2}- U)}
                                                         \notag \\
 & & \times  \, \intop_{-\infty}^{+\infty}dz \, e^{-(iy+\alpha)z^{2}+2(iy+\alpha)z\, \sqrt{\frac{J}{2N}}\sum_{i=1}^{N}S_{i}} \, .                                            \label{EGPP}
\end{eqnarray}

 Since $S_i = \{0,\pm 1\}$, it follows that
\begin{eqnarray}
\lefteqn{\sum_{\{S\}} e^{ - (iy+\alpha) \Delta\sum_{i}S_{i}^2
                                +2(iy+\alpha)z\, \sqrt{\frac{J}{2N}}\sum_{i}S_{i} }   }    \notag  \\ 
 & &   = \left[1 +e^{-(iy+\alpha) \Delta}\left(e^{2 \left(iy+\alpha\right)z\,\sqrt{\frac{J}{2N}}}+
            e^{-2(iy+\alpha)z\, \sqrt{\frac{J}{2N}}}\right)\right]^{N}                     \notag \\
 & &                                                                                       \label{B1} \\  
 & &   = \sum_{N_{0}=0}^{N}\sum_{N_{+}=0}^{N-N_{0}}\frac{N!}{N_0!N_{+}!N_{-}!} \,
           e^{-(iy+\alpha)\Delta(N-N_{0})}                                                 \notag \\
 & &   \times \,       e^{-2(iy+\alpha)z\, \sqrt{\frac{J}{2N}}(N-N_{0}-N_{+})} \,            
                    e^{ 2(iy+\alpha)z\, \sqrt{\frac{J}{2N}}N_{+}}           \, ,           \label{B2}  
\end{eqnarray}
where the last result is obtained by applying the binomial expansion twice to the result in 
Eq. (\ref{B1}).
  Now, placing the order parameters $M$ and $Q$ in Eq. (\ref{B2}) and
inserting this result into Eq. (\ref{EGPP}), one obtains
\begin{eqnarray}
 \lefteqn{ Z_{\gamma}(U,\alpha)  = \frac{1}{\sqrt{4\pi \gamma}}e^{-\alpha U}\sum_{N_{0}=0}^{N}
                                \sum_{N_{+}=0}^{N-N_{0}}\frac{N!}{N_0!N_{+}!N_{-}!}    }    \notag  \\ 
 & &   \times \,  \intop_{-\infty}^{+\infty}dy\, \left(\frac{iy+\alpha}{\pi} \right)^{1/2}
                          e^{-\frac{y^2}{4\gamma}-(iy+\alpha)(\Delta Q -U)}                          \notag  \\
 & &   \times \, \intop_{-\infty}^{+\infty} dz \, 
                          e^{-(iy+\alpha)z^2 +2(iy+\alpha)\, \sqrt{\frac{J}{2N}}\,M z}  \, . \label{IV_8}  
\end{eqnarray}
  This expression can be integrated by gaussian formulas to produce
\begin{eqnarray}
 \lefteqn{ Z_{\gamma}(U,\alpha)  =   \sum_{N_{0}=0}^{N}\sum_{N_{+}=0}^{N-N_{0}}
               \frac{N!}{N_{0}!N_{+}!N_{-}!}                                     } \notag  \\ 
    & & \times \,  e^{-\alpha (\Delta Q -\frac{J}{2N} M^2)  -  
                   \gamma(\Delta Q-U-\frac{J}{2N} M^2)^2}.                        \label{IV_9}  
\end{eqnarray}
   The solution for this ensemble notoriously brings forth the counting factor for the number of microscopic states 
corresponding to the macrostate defined by $M$ and $Q$. 
   These order parameters indeed define the energy $E$ of a configuration given by the Hamiltonian in Eq. (\ref{eq:H(S)}),
\begin{equation}
             E = \Delta Q - \frac{J}{2N} M^2  \, .       \label{eq:E_BC}
\end{equation}   
   Hence, it is convenient to write explicitly the extended partition function as a function of those order parameters, 
\begin{eqnarray}
  \lefteqn{  Z_{\gamma}(U,\alpha)                                                     } \notag  \\
& & =\, \sum_{Q=0}^{N}\sum_{M=-Q}^{M=Q}\frac{N!}{\left(N-Q\right)!\left[\frac{1}{2}\left(Q+M\right)\right]!
                                                 \left[\frac{1}{2}\left(Q-M\right)\right]!}       \notag  \\
& & \times \,  e^{-\alpha (\Delta Q-\frac{J M^{2}}{2 N})
                  -\gamma (\Delta Q-\frac{J M^{2}}{2 N}-U)^{2}}          \label{Q_BC_novo}
\end{eqnarray}

   Before studying the thermodynamic features presented by this ensemble as a function
of finite $\gamma$, it is important to show explicitly the limiting microcanonical behavior of 
this ensemble.

\subsection{Microcanonical limit and negative response functions}

  Let us firstly explore the limit $\gamma \rightarrow \infty$ to obtain the microcanonical ensemble.
  Since it is required that the extended partition function is well-behaved in this limit, the sum in $Q$ 
must converge to a dominant value for some $Q$ such that $ \Delta Q - J M^2/2N -U =0$. 
  This is nothing else than the microcanonical constraint on the energy $E$.

  Next, the thermodynamic limit  $N \rightarrow \infty$ is studied.
  Here, it is convenient to work with the intensive quantities $q= Q/N$ and $ m=M/N$.
  Let us also define $K=J/2\Delta$ and $\varepsilon = U/\Delta N$ as in  Ref. \cite{Ruffo_PRL2001}.
  Equation  (\ref{eq:E_BC}) now reads $\varepsilon = q - K m^2 $ and becomes a
constraint equation for the average energy $\varepsilon$ as $\gamma \rightarrow \infty$.
 
  For large $N$, one can evaluate the {\it microcanonical} partition function $Z(\varepsilon,\alpha)$, where 
$Z(\varepsilon,\alpha) =\lim_{\gamma \rightarrow\infty} Z_{\gamma}(\varepsilon,\alpha)$,
as a variational problem.
  To this end, we consider the saddle point solution, 
$ Z(\varepsilon,\alpha) \approx e^{- N \varphi(\varepsilon,\alpha, m)} $,
where $m$ is such that the thermodynamic potential
\begin{eqnarray}
  \lefteqn{ \varphi(\varepsilon,\alpha,m)  =  \varepsilon \alpha\Delta  }       \notag  \\  
 & & + \left[q\ln\left(\frac{\sqrt{q^{2}-m^{2}}}{2(1-q)}\right)+
            \frac{m}{2}\ln\left(\frac{q+m}{q-m}\right)+\ln\left(1-q\right)\right] ,  \notag \\ 
 & &  \label{f_NonEQ}
\end{eqnarray}
is minimized for each average energy per site $\varepsilon$.
   To obtain $\varphi(\varepsilon,\alpha, m)$, we also applied the Stirling approximation for large $N$ 
to $Z(\varepsilon,\alpha)$ in Eq. (\ref{Q_BC_novo}).
   The above expression, Eq. (\ref{f_NonEQ}), was kept as a function of $q$ and $m$, to recognize that
the term inside the brackets is the correct microcanonical entropy $- s_{\rm micro}(\varepsilon, m)$ obtained in 
\cite{Ruffo_PRL2001} as a function of the parameter mean energy $\varepsilon$ and mean magnetization $m$.
   The entropy is an even function of $m$ and a nonconcave function of the independent variables 
$m$ and $\varepsilon$, as respectively shown in Fig. 1 and 2b. 
   This fact has striking consequences for the response functions specific heat and specific susceptibility.
        
    Here, we remark that the microcanonical entropy is not always an analytic function.
    As a consequence, gaps may develop in the magnetization for some values of $\varepsilon$ as illustrated in Figure 1 for the coupling $\Delta/J= 0.462407$.
    This means that the system presents ranges of disconnected magnetization as function of 
$\varepsilon$ and $\Delta/J$.
    Thus, we cannot move continuously from one domain of magnetization to any other,
leading to the so-called microcanonical ergodicity breaking \cite{Ruffo_PRL2005},
which is not related to any phase transition.
    The condition for unaccessible magnetization states can be easily determined from the expression for the entropy. 
     Thus, one finds that those gaps start at $\varepsilon = \Delta/2J$ and increase as the energy $\varepsilon$ is lowered.


\begin{figure}[!t]
\begin{center}
\includegraphics[width=0.43\textwidth]{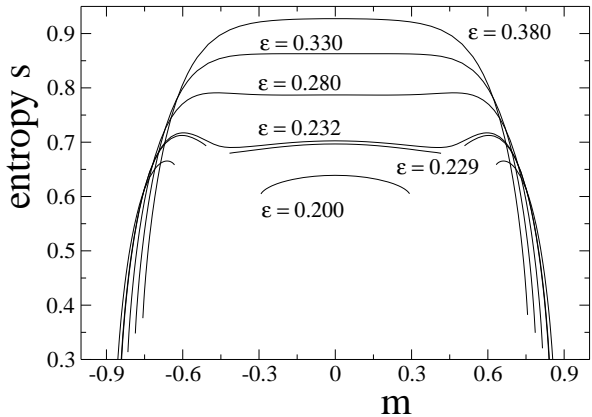}
\caption{Entropy $s_{\rm micro}(\varepsilon,m)$ for some values of $\varepsilon$ with $\Delta/J = 0.462407$. Gaps in the magnetization correspond to unaccessible states.} 
\label{fig:1}
\end{center}
\end{figure}

   Now, back to the EGE approach, it is worthy of mention that we are not evaluating a Laplace integral of the usual canonical partition function:
the extended thermodynamic potential per site $\varphi_{\gamma}$ results from a modified partition function
$Z_{\gamma}$,
\begin{equation}
   \varphi_{\gamma}(\varepsilon,\alpha)=               
                -\lim_{N\rightarrow\infty}\frac{1}{N} \ln Z_{\gamma}(\varepsilon,\alpha) \, ,
\end{equation}
 which transforms nonequilibrium states of the canonical ensemble into equilibrium states of the extended
ensemble.
  Here, the dependence of $\varphi_{\gamma}$ on $m$ has been omitted because we are already assuming that the minimization
in $m$ has been accomplished.
  As emphasized, the nonequivalence of ensembles (microcanonical and canonical) is a consequence of
the anomalous behavior of the microcanonical entropy characterized by the existence of convex parts in 
$s_{\rm micro}(\varepsilon, m)$.
  The nonconcavity of the entropy function means that the system contains several energy-dependent equilibrium states, revealed in the microcanonical ensemble, which do not have their counterpart in the temperature-dependent equilibrium
states in the canonical description.
  Thus, the new term in $\gamma$ turns such points $\varepsilon(T)$ into equilibrium points in the extended ensemble
\cite{Touchette_JDP119,Touchette_E73,Touchette_E74,Touchette_A365}.
  The usual thermodynamic potential $\varphi(\varepsilon,\alpha)$ is given by the minimization procedure
\begin{equation}
  \varphi(\varepsilon,\alpha) = \underset{m}{\rm min}  \, \varphi(\varepsilon,\alpha,m) \, ,
\end{equation}
where the dependence of $\varphi$ on $\varepsilon$ is always kept to show that $\varphi(\alpha)$ is
calculated at the equilibrium value that minimizes this potential.

\begin{figure}[!t]
\begin{center}
\includegraphics[width=0.42\textwidth]{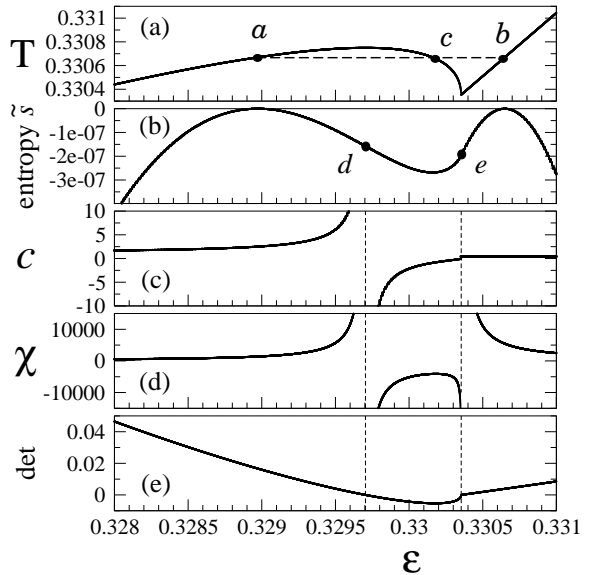}
\caption{Microcanonical behavior of the BC model with $\Delta/J = 0.462407$.
  (a) Microcanonical temperature as a function of the average energy $\varepsilon$.
      The horizontal dashed line corresponds to the canonical critical temperature of the transition.
  (b) The shifted microcanonical entropy $\tilde{s}(\varepsilon) =  s_{\rm micro}(\varepsilon)- (A + B\varepsilon)$.
  The subtraction is performed to visualize the nonconcavity of the entropy in relation to the linear
  function joining $s_{\rm micro}(\varepsilon_a)$ to $s_{\rm micro}(\varepsilon_b)$.
  (c) Specific heat $c(\varepsilon)$. It presents two poles located by the zeros of the determinant
  $d(\varepsilon,m)$, where $m$ stands for the values that maximize the entropy at $\varepsilon$.
  Those poles can also be read from $T(\varepsilon)$ behavior in (a).
  $c(\varepsilon)$ becomes negative in between those poles.
  (d) Specific susceptibility $\chi(\varepsilon)$. It presents two poles, again placed at
  the zeros of $d(\varepsilon,m)$ and becomes negative between them.
  (e) Behavior of the determinant $d(\varepsilon,m)$ as a function of $\varepsilon$.
   The vertical dashed lines signal the zeros of $d(\varepsilon,m)$.
  }
\label{fig:2}
\end{center}
\end{figure}


   In the present ensemble, the LF transform (\ref{eq:LF}) of 
$\varphi_{\gamma}(\varepsilon,\alpha,m)$, where $\varepsilon, \alpha$ and $m$
are independent variables, produces the correct $s_{\rm micro}(\varepsilon)$ as follows,   
\begin{equation}
  s_{\rm micro}(\varepsilon) =  \underset{\alpha}{\rm min} \, \underset{m}{\rm max} \,  
                      \{ \lim_{\gamma \rightarrow\infty} s_{\gamma}(\varepsilon,\alpha,m) \}  \, ,
\end{equation}
where $\lim_{\gamma \rightarrow\infty} s_{\gamma}$ stands for 
$\varepsilon\alpha\Delta - \varphi(\varepsilon,\alpha,m)$ in this model. 
  From this result one recovers the known thermodynamic behavior.
  Figure 2 contains our calculations for the microcanonical temperature $T(\varepsilon)$, shifted entropy
$\tilde{s}(\varepsilon) = s_{\rm micro}(\varepsilon)-(A+B\varepsilon)$, specific heat, susceptibility 
and determinant of the curvature of $s_{\rm micro}$, as a function of $\varepsilon$ for $\Delta/J = 0.462407$.
  This value of $\Delta/J$ is in the canonical first-order phase transition region but it 
is in the microcanonical second-order phase transition region.

  Since $\varepsilon = U/\Delta N$, one obtains
\begin{equation}
   \frac{1}{T(\varepsilon)} = \frac{1}{\Delta} \frac{\partial s_{\rm micro} }{\partial \varepsilon}
                              \equiv \beta(\varepsilon)                \, .
\end{equation}
  The horizontal dashed line in Fig. 2a indicates the temperature $T_{\rm can} \simeq 0.330666$ obtained by canonical methods \cite{beg_1971,beg_1974}. 
  It connects the point $T(\varepsilon_a)$ to $T(\varepsilon_b)$,
where $\varepsilon_a \simeq 0.328959$ and  $\varepsilon_b \simeq 0.330646$, are read from Fig. 2a.
  The width $\delta \varepsilon = 0.001687$ is the specific latent heat of the first-order phase transition
seen in the canonical ensemble.
  
   In Fig. 2b it is shown $s_{\rm micro}(\varepsilon)$ shifted by the {\it canonical} entropy
$s(\varepsilon) = A + B\varepsilon$, where $A \simeq 0.401447$ and $B \simeq 1.398397$ are such that 
$s(\varepsilon_a)= s_{\rm micro}(\varepsilon_a)$ and
$s(\varepsilon_b)= s_{\rm micro}(\varepsilon_b)$.
   This subtraction allows one to highlight the so called convex intruder in the specific entropy \cite{Gross_report,Gross_2000}.

  The point $c$, as signaled in Fig. 2a, corresponds to the energy $\varepsilon_c$ where occurs the minimum of the shifted entropy.
  Points $d$ and $e$ in Fig. 2b signal the energy range $( \varepsilon_d, \varepsilon_e)$
where the entropy is nonconcave, 
$\varepsilon_d \simeq 0.3297040$ and $\varepsilon_e \simeq 0.3303532$.
In Fig. 2a, we have the corresponding temperature $T(\varepsilon_d) \simeq 0.33074967$ as the maximum temperature in that energy range.

  Figure 2c shows the specific heat
\begin{equation}
  c(\varepsilon) =  \frac{d \varepsilon}{d T(\varepsilon)} 
                 = -\left. \frac{s_{mm}}{T^{2}d(\varepsilon,m)}  \right|_m \, ,                   \label{cv-micro}
\end{equation} 
where $d(\varepsilon,m)$ is the determinant of the curvature of $s_{\rm micro}(\varepsilon)$,
\begin{equation}
d(\varepsilon,m) =  
 \frac{1}{\Delta^2} \det\left(\begin{array}{cc}
  s_{\varepsilon \varepsilon}  & s_{\varepsilon m} \\
  s_{m \varepsilon}            & s_{m m}  
\end{array}\right)       \, ,
\end{equation}
where the notations $s_{\varepsilon \varepsilon}$, $s_{\varepsilon m}$  and $s_{m m}$ refer respectively to the second derivatives
$\partial^2 s_{\rm micro} /\partial \varepsilon^2$, $\partial^2 s_{\rm micro}/ \partial \varepsilon \partial m$
and  $ \partial^2 s_{\rm micro}/\partial m^2$.
   This determinant addresses the stability conditions around the stationary points $m$ and $\varepsilon$ \cite{Gross_book,Ruffo_Lecture2002}.

   Figure 2d presents the corresponding magnetic susceptibility
\begin{equation}
 \chi_{\rm micro}(\varepsilon,m) = - \frac{s_{\varepsilon \varepsilon}/\Delta^2 }{d(\varepsilon,m)}  \, .    
                                                                                                  \label{chi-micro}
\end{equation}
   The nonconcavity of the microcanonical entropy in $\varepsilon$ and $m$ renders a negative region for the
specific heat and magnetic susceptibility.
   Finally, Fig. 2e depicts the behavior of the determinant $d(\varepsilon,m)$ as a function of $\varepsilon$
with $m$ evaluated at the microcanonical equilibrium condition.
   The zeros of this determinant indicate the region where the response functions attain negative values.
   Their locations are represented by vertical dashed lines.
   Those negative values for the canonical observables happens inside the convex region related to the phase separation in the first-order thermodynamic transition.

\subsection{Finite $\gamma$ and extended thermodynamic potential}

   For finite $\gamma$, one obtains different equilibrium properties.
    As we are going to show, the full equivalence with the ones in the microcanonical ensemble 
is achieved for {\it finite} $\gamma$ only for $\Delta/J$ between the canonical 
and the microcanonical tricritical points. 
    On the microcanonical first-order transition line, one needs $\gamma \rightarrow \infty $
for such full recovery of the microcanonical results.

   The analytical solution for the extended thermodynamic potential is analogously obtained following the 
procedure leading to Eq. (\ref{f_NonEQ}),
\begin{eqnarray}
\lefteqn{  \varphi_{\gamma}(\varepsilon,\alpha,m,q)     =                       }             \notag  \\
  & &     q \ln\left(\frac{\sqrt{q^2-m^2}}{2(1-q)}\right)
         +\frac{m}{2} \ln\left(\frac{q+m}{q-m}\right) + \ln(1-q)                         \notag  \\ 
  & &    + \gamma \Delta^2 (q- Km^2-\varepsilon)^2 + \alpha \Delta(q-Km^2)  \, .   \label{phi_gamma}
\end{eqnarray}
    The basic problem concerning nonequivalent ensembles is that the true $s_{\rm micro}$ cannot be
obtained as an LF transform of the free energy $\varphi(\beta)$.
    Here, the application of the extended LF transform to $\varphi_{\gamma}$ yields the extended entropy $s_{\gamma}$, which can be read from Eq. (\ref{phi_gamma}),
   $s_{\gamma}=  \alpha \Delta(q-Km^2) - \varphi_{\gamma}$.
    This entropy is now a concave function of $\varepsilon$.
    The extended inverse temperature reads 
$\alpha=({\partial s_{\gamma}}/{\partial \varepsilon})/\Delta $.
    It characterizes stationary points analogous to the physical inverse temperature and is $\gamma$ dependent.

    Now, let us evaluate the equilibrium points of $ \varphi_{\gamma}(\varepsilon,\alpha,m,q)$.
    Notice that the microcanonical constraint for the specific quantities is not enforced here, the
variables $\varepsilon,m$ and $q$ are treated as independent variables.
    The linear term in $\gamma$ can be seen as a constrained equation, leading to the
microcanonical ensemble only for $\gamma \rightarrow \infty$.

    It was verified that all solutions of 
$   \partial \varphi_{\gamma}/ {\partial \alpha} = \varepsilon \Delta$,
as expressed by Eq. (\ref{eq:U}), and
${\partial \varphi_{\gamma}}/ {\partial m} = 0 $,
${\partial \varphi_{\gamma}}/ {\partial q} = 0 $,
for a fixed $\varepsilon$, are only the ones given, for example, for $T(\varepsilon)$ in Fig. 2a.
    However, those solutions are not stable for all $\gamma$.
    Since the analytical expression $\varphi_{\gamma}$ comes from the saddle-point approximation in
Eq. (\ref{Q_BC_novo}), one needs to study the stability of those EGE solutions as a function of $m$ and $q$.
    To this end, the determinant of the Hessian matrix,
\[
 d(m,q)
    =\det\left(\begin{array}{cc}
\frac{\partial^{2}\varphi_{\gamma}}{\partial m^{2}} & \frac{\partial^{2}\varphi_{\gamma}}{\partial m\partial q}\\
\frac{\partial^{2}\varphi_{\gamma}}{\partial q\partial m} & \frac{\partial^{2}\varphi_{\gamma}}{\partial q^{2}}
\end{array}\right) \, ,
\]
is analyzed in the $T$ versus $\varepsilon$ plane as a function of $\gamma$.
    This amounts to exploring which points $\{m,q\}$ minimize $\varphi_{\gamma}$ for fixed $T$ and
$\varepsilon$, and satisfy the condition $d(m,q) \geq 0$.

\begin{figure}[!t]
\begin{center}
\includegraphics[width=0.43\textwidth]{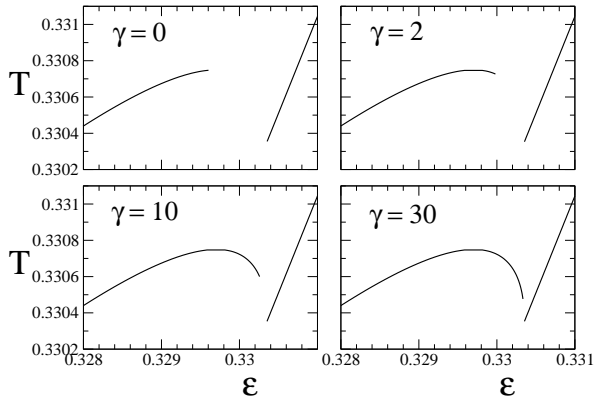}
\caption{EGE temperatures for some values of $\gamma$ with $\Delta/J = 0.462407$.} 
\label{fig:3}
\end{center}
\end{figure}


\begin{figure}[!t]
\begin{center}
\includegraphics[width=0.43\textwidth]{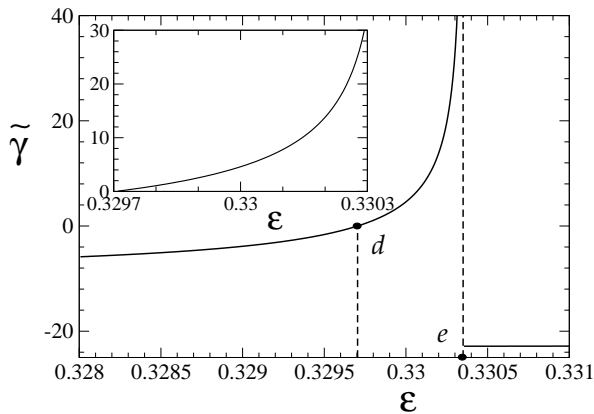}
\caption{$\tilde \gamma \equiv -1/2\Delta T^2(\varepsilon) c(\varepsilon)$ presents positive values
for $\varepsilon \in (\varepsilon_d, \varepsilon_e)$: energy range of nonconcave entropy.
 Negative values of $\tilde \gamma$ occur at energies out of this range. Here $\Delta/J = 0.462407$.} 
\label{fig:4}
\end{center}
\end{figure}


  Figure 3 shows the lines of stable points for different values of $\gamma$.
  Notice that the figure for $\gamma=0$ corresponds to the canonical results, but we have not 
included the Maxwell construction. 
  All presented states are the stable ones in the canonical ensemble.
  This procedure selects solutions $T(\varepsilon)$ for energies where the entropy is a concave function.
  The gap  in $\varepsilon$  corresponds to the region where one observes negative values in the specific heat and in the susceptibility.
   As $\gamma$ increases, one recovers the microcanonical solution. 
   In fact, for sufficiently large $\gamma$, $s_{\gamma}(\varepsilon)$
becomes entirely concave and continuous on 
$\varepsilon \in (\varepsilon_d, \varepsilon_e)$ \cite{Touchette_E73,Costeniuc07},
\begin{equation}
   \frac{\partial^2 s_{\gamma}}{\partial \varepsilon^2} < 0  \, .     \label{gammaP}     
\end{equation}

    The addition of the term in $\gamma$ to the usual Legendre transform changes the energy range
where the nonconcavity of the $\it canonical$ entropy is observed.
    How this energy range is reduced as $\gamma$ increases can be easily evaluated from Eq. (\ref{gammaP}).
    This implies the following condition on $\gamma$, 
\begin{equation}
   \gamma >  \frac{-1}{2\Delta  T^2(\varepsilon) c(\varepsilon)} \, .
\end{equation}
     But in view of the specific heat $c(\varepsilon) < 0$ for $\varepsilon \in (\varepsilon_d, \varepsilon_e)$,
 one obtains $\gamma > 0$ in this range, as exhibited in the inset of Fig. 4.
     Figure 4 shows the behavior of  $\tilde{\gamma} \equiv -1/2 \Delta  T^2(\varepsilon) c(\varepsilon)$
for energies out of that range, too.
     This figure highlights the minimum value of $\gamma$ to achieve equivalence with the
microcanonical ensemble for  $\varepsilon \in (\varepsilon_d, \varepsilon_e)$.
     The full equivalence in this energy range is reached when  $\gamma \simeq 4950$
for the example with coupling $\Delta/J=0.462407$.
     Negative values for $\gamma$ have been considered in \cite{Morishita_07} to enhance Monte Carlo sampling.
     Here, a negative $\gamma$ converts microcanonical stable states at $\varepsilon$ to unstable ones when $\varepsilon < \varepsilon_a$ or $\varepsilon > \varepsilon_b$. 
     Figure 5 shows, for all values of $\Delta/J$ between the canonical and microcanonical tricritical points,
the minimum $\gamma$ needed to recover the exact microcanonical solution.
     From the canonical approach, a first-order phase transition starts at   $\Delta/J \simeq 0.46209812$,
but from a microcanonical analysis the true first-order transition starts at $\Delta/J \simeq 0.46240788$.
    The EGE approach distinguishes those transition regions presenting finite values for $\gamma$, to recover the full thermodynamic features of this model when $\Delta/J$ is between those tricritical points.

\begin{figure}[!th]
\begin{center}
\includegraphics[width=0.43\textwidth]{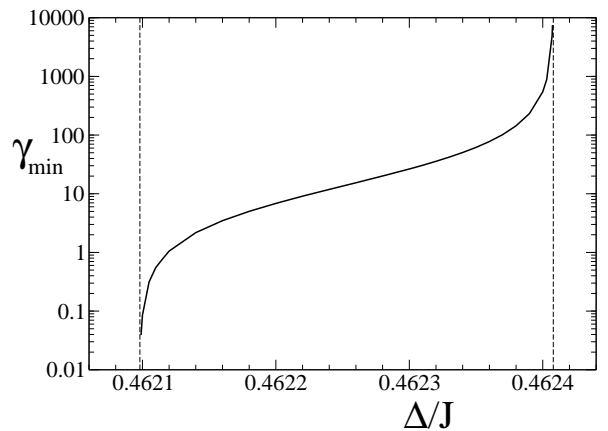}
\caption{Minimum $\gamma$ needed to recover the full microcanonical solution. The vertical dashed lines signal
the canonical ($\Delta/J \simeq 0.46209812$) and  microcanonical ($\Delta/J \simeq 0.46240788$) tricritical points.} 
\label{fig:5}
\end{center}
\end{figure}


\section{Conclusions}

   In conclusion, the analysis of the BC model shows how the stable states present in the
microcanonical approach but not found in the canonical one can be obtained from EGE.
   This approach leads to analytical expressions for the extended free energy and entropy in a simple way, and quantifies the nonequivalence of ensembles between the tricritical points.
   The EGE formulation exhibits negative specific heat, like the microcanonical one, in the canonical
first-order phase transition region. 
   This also happens between the tricritical points where $\gamma$ is finite.
   As a consequence, this remark may open a way of finding tricritical points in systems where 
analytical solutions can not be obtained. 
    Thus, an appropriated Monte Carlo method based on EGE should be preferable than the
standard one where sampling relies on the Boltzmann weight.

\section{Acknowledgement}

     The authors acknowledge support by FAPESP and CAPES (Brazil).

\end{document}